\documentclass[a4paper,11pt,headings=big,DIV=12]{scrartcl}
\usepackage{amsmath,amssymb}
\usepackage{color,xcolor}
\definecolor{blue}{rgb}{0,0,0.5} 
\usepackage[colorlinks,linkcolor=blue,urlcolor=blue,citecolor=blue]{hyperref}
\usepackage{graphicx}
\addtokomafont{disposition}{\rmfamily\boldmath}
\usepackage[utf8]{inputenc}
\usepackage{slashed}
\usepackage{multirow}
 \usepackage{bbold}
 \usepackage{mathbbol}
 
\usepackage[utf8]{inputenc}
\allowdisplaybreaks
\newcommand{\ONE}{\mathbb{1}}

\newcommand{\vev}[1]{\langle #1 \rangle}

\newcommand{\Tr}{\text{Tr}}

\newcommand{\Rtr}[2]{\Theta}

\newcommand*{\mathcolor}{}
\def\mathcolor#1#{\mathcoloraux{#1}}
\newcommand*{\mathcoloraux}[3]{%
  \protect\leavevmode
  \begingroup
    \color#1{#2}#3%
  \endgroup
}

\begin{document}

 \thispagestyle{empty}

\begin{flushright}
\begin{tabular}{l}
UUITP-4/19\\
\end{tabular}
\end{flushright}
\vskip1.5cm

\begin{center}
{\Large\bfseries \boldmath Boundary gauge and gravitational anomalies from Ward identities}\\[0.8 cm]
{\Large%
Vladimir Prochazka
\\[0.5 cm]
\small
 Department of Physics and Astronomy, Uppsala University,\\
Box 516, SE-75120, Uppsala, Sweden 
} \\[0.5 cm]
\small
E-Mail:
\texttt{\href{mailto:vladimir.prochazka@physics.uu.se}{vladimir.prochazka@physics.uu.se}}
\end{center}

\bigskip

\pagestyle{empty}
\begin{abstract}

We consider the two-point functions of conserved bulk currents and energy-momentum tensor in a boundary CFT defined on $\mathbb{R}_-^{1,2}$. Starting from the consistent forms of boundary gauge and gravitational anomalies we derive their respective contributions to the correlation functions in the form of anomalous Ward identities. Using the recently developed momentum space formalism we find an anomalous solution to each of these identities depending on a single undetermined form-factor. We study the solution in two different kinematic limits corresponding to small and large momentum $p_n$, perpendicular to the boundary. We find that the anomalous term interpolates between a non-local form resembling the standard anomaly-induced term in a two-dimensional CFT at small $p_n$ and Chern-Simons contact terms at large $p_n$. Using this we derive some consistency conditions regarding the dependence of these anomalies on the boundary conditions and discuss possible cancellation mechanisms. These ideas are then demonstrated on the explicit example of free, massless three-dimensional fermion. In particular we manage to obtain the respective anomalies via a diagrammatic  momentum space computation and expose the well-known relation between bulk parity anomaly and boundary gauge anomalies. 

\end{abstract}

\newpage 

\setcounter{tocdepth}{3}
\setcounter{page}{1}
\tableofcontents
\pagestyle{plain}

\section{Introduction}
\subsection{From anomalies to Chern-Simons}
The intimate connection between quantum anomalies and correlators of conserved currents has a long history. 
In two-dimensional CFTs there are 't Hooft anomalies (by that we mean anomalies of continuous, global symmetries) that arise for example when the theory includes chiral fermions.
 The most straight-forward way to understand and study them is through the anomalous Ward identities in momentum space \cite{Adam:1993fy,Bertlmann:2000da}. More specifically chiral and gravitational anomalies appear as contact terms in the conservation equations 
\begin{eqnarray} \label{eq:2dAnom}\notag
p_\mu \vev{j^\mu(p)j^\nu(-p)} &\sim & \epsilon^{\mu \nu} p_\mu \; , \\
p_\mu \vev{T^{\mu \nu}(p)T^{\rho \sigma}(-p)} &\sim & \epsilon^{\mu \nu} p_\mu (p^\rho p^\sigma - \eta_{\rho \sigma} p^2) \; ,
\end{eqnarray}
where $j^\mu$ is a conserved current (E.g. left/right handed chiral current) and $T^{\mu \nu}$ is the energy-momentum tensor of the theory. In \cite{AlvarezGaume:1983ig} it was shown that the presence of gravitational anomaly severely restricts the form of energy-momentum tensor correlators and therefore leads to observable physical effects. \\
If we consider three-dimensional theories, by contrast, there are no 't Hooft anomalies and instead one encounters the parity anomaly. The latter can be also related to two-point functions of conserved currents through the contact terms arising from background Chern-Simons (CS) counterterms \cite{Closset:2012vp}
\begin{eqnarray}\label{eq:3dAnom} \notag
\vev{j^\mu(p)j^\nu(-p)} &\ni & \epsilon^{\mu \nu \rho}p_\rho \; , \\
\vev{T^{\mu \nu}(p)T^{\rho \sigma}(-p)} &\ni & (\epsilon^{\mu \rho \lambda}p_\lambda(p^\nu p^\sigma- \eta^{\nu \sigma})+(\mu \leftrightarrow \nu))+(\rho \leftrightarrow \sigma) \; .
\end{eqnarray}
In this work it was shown that in a certain normalization the fractional part of the coefficient of these contact terms is a physical quantity for the given CFT.
This result provided a systematic tool for understanding various phases and dualities of $(2+1)-$dimensional parity violating theories (see for example \cite{Seiberg:2016gmd,Komargodski:2017keh}).\\
  The connection between two-dimensional 't Hooft anomalies and the CS can be made if we place a three-dimensional theory with contact terms of the type \eqref{eq:3dAnom} on a space with a boundary  \cite{Niemi:1983rq,AlvarezGaume:1984nf}. The CS terms are no longer gauge-invariant, with the respective variations being proportional to gauge and gravitational anomalies at the boundary. In the case of theory with integer CS terms, their anomalous gauge variation can be canceled by a contribution from boundary degrees of freedom. This is known as the Callan-Harvey mechanism \cite{Callan:1984sa}. \\
 As an example we can consider a theory with massive bulk fermions, which is empty in the IR and therefore leads to integer CS terms.  On the other hand, by imposing suitable boundary conditions (see for example \cite{Aitken:2017nfd,Dimofte:2017tpi}) this theory contains massless chiral edge modes whose (integer) anomaly exactly cancels the CS variation. Notice that in this case the anomaly inflow describes a system of decoupled CS in the bulk and boundary degrees of freedom. 
 
\subsection{Set-up of the paper}

 In the present work we will consider a more general situation where the anomaly inflow picture described above doesn't apply. In particular we will look at a system with propagating, ungapped degrees of freedom in the (2+1)-dimensional bulk. These degrees of freedom will be subject to boundary conditions that induce nontrivial dynamics at the edge. We will show that gauge anomalies arise that cannot be removed by neither bulk nor boundary local counterterms and provide a general framework to compute them from the bulk correlators.
 Moreover we will see that the anomaly implies an appearance of a massless pole in the current correlators which we  
associate with the emergence of chiral edge currents. The emergent boundary currents remind one of the quantum Hall effect but we stress that we don't quite reproduce all the physics of quantum Hall systems described in the literature since the bulk is not topological. In this sense the present work goes beyond the known results of \cite{Fradkin:1991wy,Blok:1990mc}.\\
 More concretely we will start from a general set-up of (2+1)-dimensional CFT with some conserved currents  corresponding to the (global) symmetries of the theory. We will then introduce a (1+1)-dimensional boundary, which turns this theory into a b(oundary)CFT (provided certain boundary conditions are satisfied). If we then couple the currents to background gauge fields, the respective conservation laws will be broken by terms supported at the boundary. Given the fact that both the (1+1)-dimensional anomalies \eqref{eq:2dAnom} and (2+1)-dimensional parity anomalies \eqref{eq:3dAnom}  depend on the two-point functions of the related quantities, we expect this to be the case for boundary anomalies too.\footnote{For example it was shown recently that the boundary conformal anomalies can be fully determined from the two-point functions of energy-momentum tensor \cite{Bianchi:2015liz,Herzog:2017kkj}.} We will determine this relation precisely and identify the related anomalous structures. \\
Given the contact term nature of the anomalies it is natural to want to work in momentum space, which is sensitive to the coincident behaviour of the correlators. Indeed it was shown recently in \cite{Bzowski:2018fql} that a wealth of information about various anomalies that could not be accessed by other methods can be obtained by solving the momentum space Ward identities.
The study of parity-even bCFT anomalies in momentum space was initiated in \cite{Prochazka:2018bpb}. In the latter work it was shown that considering correlators with explicitly non-conserved perpendicular momentum allows for new kinds of contact terms that precisely account for boundary conformal anomalies. Here we will extend this work by including parity-odd gauge and gravitational anomalies.\\  
 The main original contribution of this paper is the solution of anomalous Ward identities \eqref{eq:WImomentumCurr}, \eqref{eq:WImomentumTen} given by the equations \eqref{eq:CurrWIgenSol} and \eqref{eq:TenWIgenSol} respectively. As a side result we obtain criteria \eqref{eq:Jcondition},\eqref{eq:Tcondition} for the class of theories studied in this paper. Furthermore our massless free fermion results \eqref{eq:CurrExample}, \eqref{eq:Tenexample} are novel to the author's knowledge. \\
 The present work is structured as follows. In Section \ref{sec:Anomalies} we set up the formalism of the paper by discussing the boundary conditions for the currents and energy-momentum tensor and their practical implementation. Here we also introduce the relevant anomalies and the respective anomalous Ward identities in position space. Section \ref{sec:Momentum} contains most of the novel results of this paper. In it we explore the consequences of anomalous Ward identities in the momentum space and find their particular anomalous solution. In Section \ref{sec:FreeFerm} we will present the computation of U(1) and gravitational anomalies for free, massless (2+1)-dimensional fermion and finally we conclude in Section \ref{sec:Conclusions}.

\section{Boundary anomalies} \label{sec:Anomalies}
\subsection{Conserved charges and boundary conditions}
Before we start let us introduce the notational conventions used in this paper.
We will consider a (2+1)-dimensional theory with a Lorenzian signature $(+,-,-)$ containing a planar boundary placed at $x^n=0$ (for a spacelike direction $x^n$). The coordinate system will be denoted by $x^\mu=(x^i,x^n)$ with Roman letters denoting the (1+1)-dimensional coordinates parallel to the boundary. Throughout the paper we will use the  vector notation for the momenta parallel to the boundary $\underline{p} \leftrightarrow p_{i}$. \\
Let us now assume that the theory at hand has a global bulk symmetry corresponding to a conserved current $j^\mu$ 
\begin{equation} \label{eq:jcons}
\partial_\mu j^\mu = 0 \; .
\end{equation}
We would now like to construct a corresponding conserved charge. The bulk Noether charge will satisfy
\begin{equation}
\partial_0 Q= \partial_0\int dx^1 dx^n j^0=- \int dx^1 dx^n \partial_n j^n = -\int_{x^n=0}dx^1 j^n \; ,
\end{equation}
where we used the conservation of $j^\mu$ and integration by parts. This charge is conserved if
\begin{equation} \label{eq:CurrBC1}
j^n |_{x^n=0}= \partial_i \tilde{j}^i \; ,
\end{equation}
for some local, gauge-invariant boundary current $\tilde{j}^i$ with scaling dimension $\Delta_{\tilde{j}}=1$. Since the anomalous dimension of boundary current $\tilde{j}^i$ vanishes, it should be conserved in a unitary bCFT and we arrive at a simpler boundary condition 
\begin{equation} \label{eq:CurrBC2}
j^n |_{x^n=0}= 0 \; ,
\end{equation}
which is what we will assume in the present work. \\
If the current satisfies \eqref{eq:CurrBC1} we can define a conserved charge
\begin{equation}
Q'=Q+ \int_{x^n=0} dx^1 \tilde{j}^0 \; .
\end{equation}
Note that an equivalent effect is achieved by redefining the parallel components of the current
\begin{equation} \label{eq:Jshift}
j^{'i}= j^i + \delta(x^n) \tilde{j}^i \;
\end{equation}
For example \eqref{eq:CurrBC1} is trivially satisfied if $\tilde{j}^i$ is a conserved boundary current, which can happen in the presence of boundary degrees of freedom charged under the symmetry group. \\
Next we will look at the bulk energy-momentum tensor $T^{\mu \nu}$. The translational invariance in directions parallel to the boundary remains preserved, which yields the two conserved currents 
\begin{equation} \label{eq:EMTcons}
\partial_\mu T^{\mu i}=0 \; .
\end{equation}
Just as above, the corresponding bulk charges will satisfy
\begin{equation}
\partial_0 Q^i= \partial_0\int dx^1 dx^n T^{0i}= -\int dx^1 dx^n \partial_n T^{ni} = -\int_{x^n=0}dx^1 T^{ni} \; .
\end{equation}
Hence a conserved charge can be defined if
\begin{equation} \label{eq:EMTBC1}
T^{ni} |_{x^n=0}= \partial_j \tilde{t}^{ji} \; ,
\end{equation}
for some local boundary operator $\tilde{t}^{ji}$ with the classical scaling dimension $2$.\footnote{In general, for coupled theories with boundary RG flows, the operator $\tilde{t}^{ij}$ can acquire anomalous dimension as seen in \cite{DiPietro:2019hqe}. Nevertheless at a conformal fixed point in (2+1) dimensions, which is the case considered in this paper, the anomalous dimension and l.h.s. of \eqref{eq:EMTBC1} have to vanish (see \cite{Nakayama:2012ed} for a more detailed discussion of constraints imposed by the boundary conformal invariance).  }
As before the conserved charges are obtained via a shift
\begin{equation}
Q^{'i}=Q^i+ \int_{x^n=0} dx^1 t^{0i} \; ,
\end{equation}
which is equivalent to the redefinition of the parallel components
\begin{equation} \label{eq:bTensor}
T^{'i j}= T^{i j}+ \tilde{t}^{ij} \delta(x^n)
\end{equation}
Including the boundary breaks the Lorentz group down to the single generator subgroup $SO(1,1)$. By computing the corresponding Noether's charge we find that its conservation requires $\tilde{t}^{ij}$ to be symmetric. \\
In addition the theory can be invariant under the boundary conformal group which also contains dilatations and the two preserved special conformal transformations. By computing the corresponding charges and assuming \eqref{eq:EMTBC1} it can be shown that the operator $\tilde{t}_{ij}$ has to be traceless ($\tilde{t}_i^i=0$) and conserved ($\partial_i\tilde{t}^{ij}=0$) by scale and special conformal invariance respectively. This means that in a conformal theory $\tilde{t}_{ij}$ can be thought of as an energy-momentum tensor of some boundary degrees of freedom.\footnote{These boundary degrees of freedom can either be a decoupled 2d CFT existing at the boundary or more generally the boundary fields could also be coupled to the bulk through a fixed point coupling as in \cite{Herzog:2018lqz}.}
  The conformal invariance is only guaranteed subject to the well known Cardy condition \cite{Cardy:1984bb}
\begin{equation} \label{eq:Cardy}
T^{ni} |_{x^n=0}=0 \; .
\end{equation}
  Throughout the majority of this paper we will focus on theories possessing a conserved, traceless bulk energy-momentum tensor satisfying \eqref{eq:Cardy} with $\tilde{t}_{ij}=0$ for simplicity (boundary degrees of freedom can always be included into the present analysis via \eqref{eq:Jshift} and \eqref{eq:bTensor}) . 

\subsection{Background fields and variational principle}
To perform computations we will treat the background fields as sources for the relevant composite operators. More specifically we will couple $j^\mu$ and $T^{\mu \nu}$ to a background gauge field $A_\mu$ and the metric $g_{\mu \nu}$ respectively. In order to do this we need to specify boundary conditions for the background fields. 
The boundary conditions used in this paper will be the following
\begin{eqnarray} \label{eq:GfieldBC} \notag
\partial_n A_i|_{x^n=0}&=& 0 \\
 A_n|_{x^n=0} &=& 0 \quad ,  
\end{eqnarray}
while the parallel components $A_i|_{x^n=0} \equiv \hat{A}_i$ are not restricted.\footnote{\label{ft1}If we wish to extend the background fields into the upper half-plane the boundary conditions \eqref{eq:GfieldBC} are consistent with the classical parity transformations under the reflection through the boundary
\begin{eqnarray*} \notag
A_n &\stackrel{x^n \to - x^n}{\to}& - A_n \\
A_i &\stackrel{x^n \to - x^n}{\to}& \;  A_i \\
\end{eqnarray*}}\\
Similarly for the metric (in a suitable coordinate system) we impose
\begin{eqnarray} \label{eq:MetricBC} \notag
\partial_n g_{ij}|_{x^n=0}&=& 0 \\
 g_{in}|_{x^n=0} &=& 0 \quad ,  
\end{eqnarray}
with the parallel metric $g_{ij}|_{x^n=0} \equiv \tilde{g}_{ij}$ being unrestricted function of the boundary coordinates. 
The connected correlators will be defined through functional derivatives of the partition function w.r.t. gauge field and the metric 
\begin{equation}
j^\mu(x) \leftrightarrow \frac{1}{i}\frac{\delta}{\delta A_\mu(x)}; \quad
T^{\mu \nu}(x) \leftrightarrow 2i\frac{\delta}{\delta g_{\mu \nu}(x)} \; ,
\end{equation}
where we will set all the background fields to zero/flat value at the end. To be able to perform the functional differentiation we also need to define variational rule consistent with \eqref{eq:GfieldBC} and \eqref{eq:MetricBC}. For a field $\phi_N$ satisfying Neumann boundary condition such a variational rule can be defined (c.f. \cite{Prochazka:2018bpb}) via the method of images
\begin{equation} \label{eq:NeumannDelta}
\frac{\delta \phi_N(x)}{\delta \phi_N(y)} \equiv \delta^N(x,y) \equiv \delta (x-y) + \delta(\bar{x}-y) \; ,
\end{equation}
where $\bar{x}= (x^i,-x^n)$ is the position of image in the unphysical region. Note that by definition $\delta^N$ reduces to the usual delta function in the lower half plane and satisfies $\frac{\partial}{\partial x^n}\delta^N(x,y)|_{x^n=0}=0$.

\subsection{Anomalous Ward identities}
When coupling the theory to background fields, operator equations \eqref{eq:jcons}, \eqref{eq:EMTcons} can receive c-number corrections corresponding to quantum anomalies. Since there are no bulk anomalies of continuous symmetries in odd dimensions, these corrections therefore have to be supported at the boundary. 
To derive the form of the anomaly we will assume that the theory is classically invariant under gauge transformations and diffeomorphisms consistent with the boundary conditions \eqref{eq:GfieldBC}, \eqref{eq:MetricBC}. In particular this means that at the boundary the gauge transformations take the usual form 
\begin{equation} \label{eq:BdaryGaugeTransf}
\hat{A}_i \to \partial_i \hat{\alpha}
\end{equation}
 where the boundary background gauge field is defined through a BC $\hat{A}_i= A_i |_{x^n=0}$ and  $\hat{\alpha}$ is an arbitrary function of the parallel coordinates $x^i$.\footnote{The function $\hat{\alpha}$ can be thought of as the boundary limit of bulk gauge transformation $\delta A_{\mu}= \partial_\mu \alpha$ satisfying Neumann BC.} Since there is no bulk anomaly the anomalous gauge transformation of the generating functional has to be localized to the boundary (I.e. $\delta_{gauge} W = \int_{x^n=0} \hat{\alpha} \dots$). The form of the boundary anomaly can then be determined by solving the Wess-Zumino consistency conditions subject to \eqref{eq:BdaryGaugeTransf} (this was done for example in \cite{Bardeen:1984pm, Stone:2012ud}), which gives the following expression for the consistent unintegrated anomaly: 
\begin{equation} \label{eq:currAn}
\partial_\mu \vev{j^\mu}=  \frac{a_\partial}{4 \pi} \delta(x^n)\epsilon^{ij}\partial_i \hat{A}_j \; .
\end{equation}
 Similarly by varying the generating functional w.r.t. parallel diffeomorphisms we arrive at the consistent  form of the gravitational anomaly
\begin{equation}\label{eq:tenAn}
\nabla_\mu \vev{T_i^\mu}= \frac{e_\partial}{192 \pi} \delta(x^n)\epsilon^{kl} \partial_j  \partial_{k}{\hat{\Gamma}_{l i}}^j \; ,
\end{equation}
where $\hat{\Gamma}$ are the Christoffel symbols compatible with the boundary metric.\footnote{In the Gaussian normal coordinates, the boundary metric can be defined through the $x^n \to 0$ limit of the bulk component $g_{ij}$.}
Varying both sides \eqref{eq:currAn}, \eqref{eq:tenAn} w.r.t. $A_{j}(x')$, $g_{kl}(x')$ respectively we obtain anomalous Ward identities in the flat spacetime
\begin{equation} \label{eq:currWI1}
\partial_\mu \vev{j^\mu(x) j^i(x')}= \frac{1}{i}\frac{\delta}{\delta A_i(x')}\partial_\mu \vev{j^\mu(x)}|_{A^\mu=0}= \frac{1}{i}\frac{a_\partial}{4 \pi} \delta(x^n)\epsilon^{ij} \partial_i \delta^N (x,x')  \; ,
\end{equation}
and
\begin{equation} \label{eq:tensorWI1}
\partial_\mu \vev{T^{\mu j}(x) T^{kl}(x')}= 2i\frac{\delta}{\delta g_{kl}(x')}\nabla_\mu \vev{T^{\mu j}(x)}|_{g_{\mu \nu}=\eta_{\mu \nu}}= 2i \frac{e_\partial}{192 \pi} \delta(x^n)\epsilon^{ij} \partial_i (\partial^k \partial^l - \eta^{kl} \partial_m^2)\delta^N (x,x')  \; ,
\end{equation}
where the Neumann delta $\delta^N$ function was defined in \eqref{eq:NeumannDelta}.

\section{Momentum space analysis} \label{sec:Momentum}

Following \cite{Prochazka:2018bpb} we want to analytically extend the correlators into the upper half-plane $x^n>0$.\footnote{In presence of branch cuts in the normal direction, we can always choose the principal branch as the definition of the correlators.} The corresponding momentum space correlators are then defined via Fourier transform as usual. The goal of this section is to identify within the respective two-point functions the structures responsible for \eqref{eq:currAn} and \eqref{eq:tenAn}. 

\subsection{Current correlator} \label{sec:CurrCorr}

 We will start our analysis with two point functions of the conserved bulk current $j^\mu$ \eqref{eq:jcons} subject to boundary condition \eqref{eq:CurrBC2}. The correlators $\vev{jj}$ 
can be defined in the lower half-space so we will analytically continue them to the unphysical upper half-space region and then take the Fourier transform over the whole of $\mathbb{R}^{1,2}$
\begin{equation} \label{eq:JJcorrMom}
\vev{j^\mu(\underline{p},p_n) j^\nu(\underline{p}',p_n')}= \int d^dx e^{i px}\int d^d x' e^{i p'x'} \vev{j^\mu(x)j^\nu(x')} \;.
\end{equation}
 Following \cite{Prochazka:2018bpb} we will further take the exceptional kinematics with $p_n>0, p_n'=0$ which explicitly breaks the perpendicular momentum conservation. Physically this corresponds to an amplitude where the incoming state gets absorbed by the boundary. The correlator \eqref{eq:JJcorrMom} doesn't vanish in this kinematical limit, instead it takes a form dictated by the boundary  conformal group. Based on the invariance w.r.t. translations parallel to boundary, the momentum space correlator takes the form
\begin{equation} \label{eq:JJmomRep}
\vev{j^\mu(\underline{p},p_n)j^\nu(\underline{p}',0)}= \delta(\underline{p}+\underline{p}') \Pi^{\mu \nu}(\underline{p}, p_n) \; ,
\end{equation}
where $\Pi^{\mu \nu}$ is some covariant tensor quantity.\footnote{Note that the conserved bulk contribution proportional to $\delta(p_n)$ is excluded as long as $p_n \neq 0$.} The goal of our analysis is to identify a structure responsible for the anomaly in $\Pi^{\mu \nu}$. \\
 We first Fourier transform \eqref{eq:currWI1} with the exceptional kinematics $p_n'=0$ to get the momentum space Ward identity
\begin{eqnarray} \label{eq:WImomentumCurr} \notag
p_\mu \Pi^{\mu j}(\underline{p}, p_n)&=& -4i \frac{a_\partial}{4 \pi}  \epsilon^{ij}p_i \; , \\
p_\nu'\Pi^{i \nu}(\underline{p}, p_n)&=&-p_j \Pi^{i j}= -4i \frac{a_\partial}{4 \pi}  \epsilon^{ij}p_j \; ,
\end{eqnarray}
where we have accounted for the extra factor of $2$ coming from the normalization of the delta function on a half-line (I.e. $\int_{\mathbb{R}} d x^n \delta(x^n)=2 \int_{\mathbb{R}_-}dx^n\delta(x^n)=2$). \\
A parity-odd solution to \eqref{eq:WImomentumCurr} consistent with Lorentz subgroup $SO(1,1)$ and scale invariance can be found
\begin{eqnarray} \label{eq:CurrWIgenSol} \notag
\Pi_{\text{an.}}^{ij}(\underline{p},p_n)&=& 4 i \frac{a_\partial}{4 \pi} \left[ \frac{\epsilon^{ik}p_k p^j + \epsilon^{jk}p_k p^i}{\underline{p}^2} \left(1-F(\frac{\underline{p}^2}{p_n^2})\right)+ \epsilon^{ij}F(\frac{\underline{p}^2}{p_n^2})\right] \; ,\\
\Pi_{\text{an.}}^{in}(\underline{p}, p_n)&=& 8 i \frac{a_\partial}{4 \pi} \frac{F(\frac{\underline{p}^2}{p_n^2})}{p_n} \epsilon^{ij} p_j \; ,
\end{eqnarray}
where $\underline{p}^2 \equiv p_i p_j \eta^{ij}$ and $F$ is a model dependent form-factor function.\footnote{In general when determining $\Pi_{\text{an.}}^{ij}$, the $SO(1,1)$ covariance fixes the three form-factors multiplying the parity-odd structures $\epsilon^{ik}p_k p^j, \epsilon^{jk}p_k p^i $ and $ \epsilon^{ij} \underline{p}^2$ respectively. Using \eqref{eq:CurrWIgenSol} and the $2d$ identity $\epsilon^{ik}p_k p^j-\epsilon^{jk}p_k p^i- \epsilon^{ij} \underline{p}^2=0$ we remain with a single form-factor that has to depend on the ratio $\frac{\underline{p}^2}{p_n^2}$ by the conformal invariance.} Note that \eqref{eq:CurrWIgenSol} is only a particular solution to \eqref{eq:WImomentumCurr} in that it doesn't include the non-anomalous, parity-even part.  The solution to non-anomalous Ward identities for position space two-point functions of conserved bulk currents and energy-momentum tensor on spaces with planar boundaries was found in \cite{McAvity:1995zd} and \cite{McAvity:1993ue}.\\
Next we can consider the extremal limits of the solution \eqref{eq:CurrWIgenSol}. 
 First we consider the $p_n \to 0$ limit, keeping $\underline{p}^2$ finite, nonzero or $\frac{p_n}{|\underline{p}|} \to 0$. Physically this limit describes excitations moving parallel to the boundary. 
 For a smooth behaviour of $\Pi^{in}$ and $(i \leftrightarrow j)$ symmetry of $\Pi^{ij}$ in this limit we have to impose
\begin{equation} \label{eq:FlowP}
F(\frac{\underline{p}^2}{p_n^2}) \stackrel{p_n \to 0}{\approx} 0 + \mathcal{O} \left( \frac{p_n^2}{\underline{p}^2} \right) 
\end{equation}
 and the limit of the anomalous solution \eqref{eq:CurrWIgenSol} becomes 
\begin{eqnarray} \label{eq:HighMomAn} \notag
\Pi_{\text{an.}}^{ij}(\underline{p}, 0)&=& 4 i \frac{a_\partial}{4 \pi} \frac{\epsilon^{ik}p_k p^j + \epsilon^{jk}p_k p^i}{\underline{p}^2} \; ,\\
\Pi_{\text{an.}}^{in}(\underline{p}, 0)&=&0\; .
\end{eqnarray}
The manifest nonlocality of this solution implies it corresponds to the behaviour of the correlator at non-coincident points in the position space. Indeed, the first line of \eqref{eq:HighMomAn} looks exactly like the anomalous, parity-odd contribution to the axial current correlator in a $(1+1)$ dimensional CFT, which is supported at non-coincident points in the position space \cite{Chen:1999zh}. Physically \eqref{eq:HighMomAn} implies the existence of chiral currents at the boundary.\footnote{Note that the presence of $\frac{1}{\underline{p}^2}$ pole in the anomalous correlator indicates the appearance of boundary degrees of freedom. This is somewhat similar to the way in which the axial anomaly implies the existence of massless bound states in the spectrum of strongly coupled gauge theories \cite{Frishman:1980dq}.}  \\
Next we will examine the solution in the $p_n \to \infty$ limit or more precisely $\frac{|\underline{p}|}{p_n} \to 0$. In this limit bulk excitations with momenta parallel to the boundary 'freeze out' leaving behind some local terms in the effective action and possibly decoupled boundary degrees of freedom.\footnote{This limit is analogous to the small momentum limit that can be used to study the IR effective action in massive QFTs. In \cite{Prochazka:2018bpb} it was used to expose the near-boundary behaviour of the bulk effective action for parity-preserving theories. In the presence of anomalous boundary currents coupling to $A_i$ (I.e. if $\tilde{j}_i \neq 0$ in \eqref{eq:Jshift}) non-local terms of the type \eqref{eq:HighMomAn} survive in the $p_n \to \infty$ limit of $\Pi^{ij}$.  } 
 In particular this means that for $p_n \to \infty$ the correlators of bulk operators become local functions of the momentum. By inspecting \eqref{eq:CurrWIgenSol} we conclude that
\begin{equation}\label{eq:FhighP}
F(\frac{\underline{p}^2}{p_n^2}) \stackrel{p_n \to \infty}{\approx} 1 + \mathcal{O} \left( \frac{\underline{p}^2}{p_n^2} \right) \; .
\end{equation} 
  Therefore we find that the solution to Ward identities \eqref{eq:WImomentumCurr} for $\frac{\underline{p}^2}{p_n^2} \ll 1$ has the form
\begin{eqnarray} \label{eq:LowMomAn} \notag
\Pi_{\text{an.}}^{i j}(\underline{p}, \infty)&=& 4i \frac{a_\partial}{4 \pi}  \epsilon^{ij} + \mathcal{O}(\frac{\underline{p}^2}{p_n^2})\; , \\
\Pi_{\text{an.}}^{n j}(\underline{p}, \infty)&=&- \frac{1}{p_n} 8i \frac{a_\partial}{4 \pi}  \epsilon^{ij}p_i + \mathcal{O}(\frac{\underline{p}^2}{p_n^2})\; .
\end{eqnarray}
Note that $\Pi_{an.}^{n j}$ actually corresponds to a semilocal contact term 
\begin{equation}
\text{sgn}(-x^n) \epsilon^{ij} \partial_i \delta^N (x,x')
\end{equation}
in the position space.\footnote{This follows from the Fourier transform property $\int_{\mathbb{R}} e^{ipx}\text{sgn}(x)= 2 i \frac{1}{p}$.}
In fact one can verify directly that \eqref{eq:LowMomAn} comes from a Fourier transform of the following terms
\begin{eqnarray} \label{eq:CSct} \notag
 \hat{\Pi}_{an.}^{ij}(x,x')&=& -\frac{a_\partial}{4 \pi i}\epsilon^{ij}[ 2 \text{sgn}(-x^n) \partial_n \delta^N(x,x')+\delta(x^n) \delta^N(x,x')] \; , \\
  \hat{\Pi}_{an.}^{nj}(x,x')&=& \frac{2a_\partial}{4 \pi i} \epsilon^{ij} \text{sgn}(-x^n) \partial_i \delta^N(x,x') \; .
\end{eqnarray} 
For $x^n \leq 0$ these are exactly the contact terms obtained by variation of the Chern-Simons term
\begin{equation} \label{eq:CurrAnEff}
S_{eff}=\frac{a_\partial}{4 \pi} \int_{x^n \leq 0}  A \wedge dA \; .
\end{equation}
 The jump in sign of the Chern-Simons contact terms \eqref{eq:CSct} when crossing the boundary into unphysical region $x^n>0$ is perhaps expected due to parity violating nature of \eqref{eq:CurrAnEff} under the reflections described in Footnote \ref{ft1}.\footnote{In fact by extending the background fields into the upper half plane the contact terms \eqref{eq:CSct} can be obtained from a CS term with coefficient proportional to $\text{sgn}(-x^n)$. Chern-Simons terms with spacetime dependent coefficients changing the sign accross the inteface are known to appear for example for massive domain wall fermion theories \cite{Kaplan:2009yg}.} 
 \\
In the bulk \eqref{eq:CSct} reduce to the expressions of  \cite{Closset:2012vp}, where it was argued that the coefficient of the CS contact terms ($\text{mod} \; 1$) is a physical quantity in a given CFT. The non-integer part of $a_\partial$ is therefore a property of the bulk CFT that cannot be changed by the boundary dynamics. 
More concretely let $\text{BC}_1, \text{BC}_2$ be two sets of boundary conditions for fundamental fields of the same CFT satisfying  \eqref{eq:CurrBC1}. Since the non-integer part of $a_\partial$ is fixed by the bulk the two anomalies can only differ by an integer 
\begin{equation} \label{eq:Jcondition}
a_\partial^{\text{BC}_1}- a_\partial^{\text{BC}_2} \in \mathbb{Z}
\end{equation}
In another words, only the integer part of $a_\partial$ depends on the boundary dynamics. \\
Let us now discuss the qualitative properties of solution \eqref{eq:CurrWIgenSol}. Clearly by choosing different boundary conditions for the function $F$ we could obtain solutions with a different qualitative behaviour. 
 In fact it can be shown that for generic boundary conditions for $F$ the solution \eqref{eq:CurrWIgenSol} can be written as a linear combination of a solution satisfying \eqref{eq:FhighP} and \eqref{eq:FlowP}, a solution of the form \eqref{eq:HighMomAn} and another solution proportional to \eqref{eq:LowMomAn}. These solutions correspond to theories with decoupled boundary currents and bulk CS terms. Let us look at some of them in more detail
\begin{itemize}
\item{First let us consider a solution with $F(\infty) \neq 0$ (in the $p_n \to 0$ limit). The $\Pi^{in}$ part of this solution is not well-defined for $p_n \to 0$ and it yields antisymmetric contribution to $\Pi^{ij}$. It can be written as a linear combination of the continuous solution satisfying \eqref{eq:FhighP}, \eqref{eq:FlowP} plus $F(\infty) \times$ \eqref{eq:LowMomAn}. We can always subtract the CS part by hand, which will yield a continuous solution with the coefficient $a_\partial(1-F(\infty))$. For those reasons we expect $F(\infty) \neq 0$ to be an integer that can be absorbed in the definition of CS level. }
\item{Second we can have $F(0) \neq 1$ (in the $p_n \to \infty$ limit). Such solution can be written as a linear combination the continuous solution plus $(1-F(0)) \times$\eqref{eq:HighMomAn}. This corresponds to a physical solution containing decoupled boundary currents with anomaly $\tilde{a}_\partial= a_\partial (F(0)-1)$. We can get such solution for example by including free chiral fermions at the boundary.   }
\end{itemize}
From the above two points we see that we can always uniquely project on the continuous solution satisfying   \eqref{eq:FlowP}, \eqref{eq:FhighP}. Therefore we will assume $a_\partial$ to be the coefficient of the continuous solution throughout this paper.  This solution interpolates between chiral current-like behaviour of \eqref{eq:HighMomAn} and CS-like bulk behaviour of \eqref{eq:LowMomAn}, which is reminiscent of the quantum Hall effect \cite{Stone:1990iw} with $a_\partial$ playing the role of Hall conductance. From above arguments we expect $a_\partial$ (including the integer part) to be a physical quantity.

\subsection{Energy-momentum tensor correlator} \label{sec:EMTcorr}
We will now repeat the same analysis for the correlators of an energy-momentum tensor satisfying \eqref{eq:Cardy}. We will again extend the $\vev{TT}$ correlators to the upper half-space and Fourier transform them. Finiteness of this Fourier integral in the parity-even sector was established in \cite{Prochazka:2018bpb}. The energy-momentum tensor version of \eqref{eq:JJmomRep} reads
\begin{equation} \label{eq:TTmomRep}
\vev{T^{\mu \nu}(\underline{p},p_n)T^{\rho \sigma}(\underline{p}',0)}= \delta(\underline{p}+\underline{p}') A^{\mu \nu \rho \sigma}(\underline{p}, p_n) \; ,
\end{equation}
for a tensor quantity $A^{\mu \nu \rho \sigma}$ consistent with the boundary conformal group. Once again, we would like to identify a structure responsible for the anomaly \eqref{eq:tenAn} within  $A^{\mu \nu \rho \sigma}$.
Fourier transforming the translation Ward identities \eqref{eq:tensorWI1} together with the definition \eqref{eq:TTmomRep} we get
\begin{eqnarray} \label{eq:WImomentumTen} \notag
p_\mu A^{\mu j kl} &=&- 8 i \frac{e_\partial}{192 \pi}  \epsilon^{ij}p_i P^{kl} \; \\
p_\rho' A^{i j \rho l}&=&-p_k A^{i j k l}= -8i \frac{e_\partial}{192 \pi}  P^{ij}\epsilon^{kl}p_k \; ,
\end{eqnarray}
where we defined a transverse projector $P^{ij}= p^i p^j - \eta^{ij} \underline{p}^2$. \\
The general solution to \eqref{eq:WImomentumTen} consistent with Lorentz, scale invariance and $(i\leftrightarrow j), (k \leftrightarrow l)$ symmetry can be determined
\begin{eqnarray} \label{eq:TenWIgenSol} \notag
A_{an.}^{ij kl}(\underline{p},p_n)&=& 8 i \frac{e_\partial}{192 \pi}[1-G(\frac{\underline{p}^2}{p_n^2})]\left( \frac{(\epsilon^{im}p_m p^j+ \epsilon^{jm}p_m p^i)P^{kl} + (\epsilon^{km}p_m p^l+ \epsilon^{lm}p_m p^k)P^{ij}}{\underline{p}^2}\right) \\ \notag
& & +8i \frac{e_\partial}{384 \pi }G(\frac{\underline{p}^2}{p_n^2})\left((\epsilon^{ik}P_{jl}+(i \leftrightarrow j)) + (k \leftrightarrow l) \right)\\
A_{an.}^{in kl}(\underline{p}, p_n)&=&-\frac{1}{p_n} 8i\frac{e_\partial}{96 \pi } G(\frac{\underline{p}^2}{p_n^2}) P^{kl}  \epsilon^{ij} p_i\; .
\end{eqnarray}
By applying the same arguments as in the Section \ref{sec:CurrCorr}  we conclude that the function $G$ has the same behaviour as $F$ at the extremal limits \eqref{eq:FlowP}, \eqref{eq:FhighP}. Thus we can readily determine the behaviour of \eqref{eq:TenWIgenSol} for $p_n \to 0, \infty$. Starting with the non-local limit $p_n \to 0$ we find a solution analogous to \eqref{eq:HighMomAn}
\begin{eqnarray} \label{eq:HighMomAnT} \notag
A_{an.}^{ij kl}(\underline{p}, 0)&=& 8 i \frac{e_\partial}{192 \pi}\left( \frac{(\epsilon^{im}p_m p^j+ \epsilon^{jm}p_m p^i)P^{kl} + (\epsilon^{km}p_m p^l+ \epsilon^{lm}p_m p^k)P^{ij}}{\underline{p}^2}\right) \\
A_{an.}^{in kl}(\underline{p}, 0)&=&0\; .
\end{eqnarray}
Just as before, the expression in \eqref{eq:HighMomAnT} looks exactly like the Fourier transform of the parity-odd contribution to the EMT correlators in a $2d$ CFT at non-coincident points \cite{Bonora:2015nqa}. \\
Next we want to consider the $p_n \to \infty$ limit. Behaviour of the parity-even contribution to $\vev{TT}$ in this limit and the associated conformal anomalies were studied in \cite{Prochazka:2018bpb}. Here we will extend this work to include parity-odd contributions. An anomalous solution to \eqref{eq:WImomentumTen} analogous to \eqref{eq:LowMomAn} has a form
\begin{eqnarray} \notag
A_{an.}^{ij kl}(\underline{p}, \infty)&=& 8i \frac{e_\partial}{384 \pi }\left((\epsilon^{ik}P_{jl}+(i \leftrightarrow j)) + (k \leftrightarrow l) \right)  + \mathcal{O}(\frac{\underline{p}^2}{p_n^2}) \; ; \\
A_{an.}^{in kl}(\underline{p}, \infty)&=& -\frac{1}{p_n} 8i\frac{e_\partial}{96 \pi } P^{kl}  \epsilon^{ij} p_i  + \mathcal{O}(\frac{\underline{p}^2}{p_n^2}) \; .
\end{eqnarray}
As before above expressions correspond to semi-local contact terms in the position space
\begin{eqnarray}  \notag
 \hat{A}_{an.}^{ijkl}(x,x')&=& -\frac{e_\partial}{384 \pi i}\hat{D}^{ijkl}[ 2 \text{sgn}(-x^n)\partial_n \delta^N(x,x')+\delta(x^n)  \delta^N(x,x')] \; ; \\
 \hat{A}_{an.}^{njkl}(x,x')&=& \frac{e_\partial}{96 \pi i} \text{sgn}(-x^n) \hat{P}^{kl} \epsilon^{ij} \partial_i \delta^N(x,x') \; ,
\end{eqnarray}
where $\hat{D}^{ijkl} = (\epsilon^{ik}(\partial^j \partial^l - \eta^{jl}\partial_m^2)+(i \leftrightarrow j)) + (k \leftrightarrow l)$ and $\hat{P}^{kl}= \partial^k\partial^l- \eta^{kl} \underline{\partial}^2$. \\
As before these contact terms can be obtained from the gravitational Chern-Simons term
\begin{equation}\label{eq:TenAnEff}
S_{eff}^g= \frac{e_\partial}{192 \pi} \int_{x^n \leq 0} \Tr [\Gamma d \Gamma + \frac{2}{3} \Gamma^3] \; ,
\end{equation} 
where the trace is taken over the matrices $\Gamma_\alpha^\beta= \Gamma_{\mu \alpha}^\beta dx^\mu $, with the sign function appearing for the same reasons as explained under \eqref{eq:CurrAnEff}. Using the same logic we conclude that for two bCFTs agreeing in the bulk, differing only by boundary conditions $\text{BC}_1, \text{BC}_2$ we have
\begin{equation} \label{eq:Tcondition}
e_\partial^{\text{BC}_1}- e_\partial^{\text{BC}_2} \in \mathbb{Z} \; .
\end{equation}
So again we see that the boundary conditions determine the integer part of gravitational contact terms and the non-integer part being a physical quantity determined by the bulk CFT \cite{Closset:2012vp}. In addition we expect $e_\partial$ to be a physical quantity for the same reasons as $a_\partial$ (see the discussion below \eqref{eq:Jcondition}).

\subsection{Anomaly cancellation} \label{sec:AnCanc}
The presence of anomalies \eqref{eq:currAn}, \eqref{eq:tenAn} prevents one from gauging the global symmetries or putting the theory on curved manifolds. It is therefore important to explore the ways to cancel them. 
To illustrate the idea we will focus on the gauge anomaly \eqref{eq:WImomentumCurr} in the following, but all the arguments can be readily applied to the gravitational anomaly too. \\
If $a_\partial \in \mathbb{Z}$ we can cancel the anomaly \eqref{eq:WImomentumCurr} by adding boundary degrees of freedom. This can be done by turning on an anomalous two-dimensional edge current $\tilde{j}^i$ located at the boundary. We select the edge current so that its anomaly is precisely $-a_\partial$\footnote{For integer $a_\partial$ this doesn't violate the charge quantization condition and therefore is allowed.} and the correct physics is described by the current $j'=j+\tilde{j} \delta(x^n)$. In practice this has an effect of shifting the anomalous contribution \eqref{eq:CurrWIgenSol} by a term proportional to \eqref{eq:HighMomAn}. Doing this we arrive at an anomaly-free, parity-odd term contributing to the correlator of $j'$
\begin{eqnarray} \label{eq:CurrWNonAn2} \notag
\Pi^{'ij}(\underline{p},p_n)&=& 4 i \frac{a_\partial}{4 \pi}F(\frac{\underline{p}^2}{p_n^2}) \left[\epsilon^{ij}- \frac{\epsilon^{ik}p_k p^j + \epsilon^{jk}p_k p^i}{\underline{p}^2} \right] \; ,\\
\Pi^{'in}(\underline{p}, p_n)&=& 8 i \frac{a_\partial}{4 \pi}\frac{F(\frac{\underline{p}^2}{p_n^2})}{p_n} \epsilon^{ij} p_j \; .
\end{eqnarray}
Subject to \eqref{eq:FlowP} and \eqref{eq:FhighP} the above expression defines a finite, non-anomalous and non-local contribution to the current correlator. Physically this describes a system of propagating degrees of freedom both in the bulk and on the boundary. In that sense, \eqref{eq:CurrWNonAn2} is different from the usual inflow scenario of gapped theories, where we only get local CS terms in the bulk. \\
Now if $a_\partial \in \mathbb{Q}$, we cannot add purely two-dimensional degrees of freedom to cancel the anomaly as this would violate the charge quantization condition. Instead, we could cancel it by a \textit{bulk} TQFT.\footnote{The author would like to thank Zohar Komargodski for suggesting this option.} To illustrate this let us consider $a_\partial=\frac{1}{m}$ for integer $m \neq \pm 1$. To cancel this anomaly we could add $U(1)_{m}$ coupled to the background gauge field through the topological current (we refer the reader to Appendix C of \cite{Seiberg:2016gmd} for details of the treatment of these theories coupled to background gauge fields). Since the TQFT doesn't have any propagating degrees of freedom, its contribution to the two-point function is of the form \eqref{eq:LowMomAn} with a coefficient $-\frac{1}{m}$, which cancels the anomaly. For $F$ satisfying boundary conditions \eqref{eq:FlowP}, \eqref{eq:FhighP},  subtracting this term from \eqref{eq:CurrWIgenSol} still leads to a non-local expression that will describe a system of physical topological as well as propagating degrees of freedom in the bulk.

\section{Free fermion example} \label{sec:FreeFerm}
\begin{figure}[h]
\begin{center}
  \includegraphics[width=80mm]{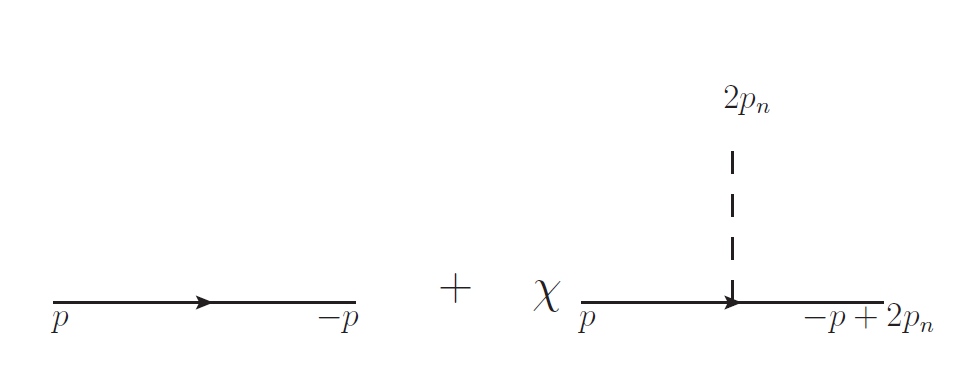}
  \caption{\small Fermion propagator in momentum space. Dashed line represents the momentum reflected from the boundary. First term corresponds to standard propagator $\frac{i \slashed{p}}{p^2}$ and the second term corresponds to the reflected propagator $\frac{i \slashed{\tilde{p}}}{p^2}$ with $\tilde{p}= (\underline{p},-p_n)$. Note that the momentum non-conservation through reflected propagator is denoted by the dashed line with $p_n \equiv(\underline{0},p_n)$.}
  \label{fig:P1}
  \end{center}
\end{figure}

In this section we will look at the example of free Dirac fermion in $2+1$ dimensions with a planar boundary placed at $x^n=0$ (we choose $x^n$ to be parallel with the Euclidean $x^3-$direction). The action for this theory will be 
\begin{equation}\label{eq:FreeFermAction}
S= \frac{i}{2}\int_{x^n \leq 0} d^3 x \sqrt{g} \bar{\psi} \overset{\leftrightarrow}{\slashed{D}} \psi \quad ,
\end{equation}
where $D_\mu$ is the Dirac operator and $\overset{\leftrightarrow}{\slashed{D}}= \overset{\rightarrow}{\slashed{D}}- \overset{\leftarrow}{\slashed{D}}$. This theory has a conserved $U(1)$ current
\begin{equation}\label{eq:FreeFermJ}
j_\mu = \bar{\psi} \gamma_\mu \psi 
\end{equation}
and energy-momentum tensor
\begin{equation} \label{eq:FreeFermT}
T_{\mu \nu} = \frac{i}{4} \bar{\psi} ( \gamma_\mu \overset{\leftrightarrow}{\partial}_\nu+\gamma_\mu \overset{\leftrightarrow}{\partial}_\nu - 2 \eta_{\mu \nu}  \overset{\leftrightarrow}{\slashed{D}}) \psi \; .
\end{equation}
A standard way to define boundary conditions for fermions is to use projectors \cite{Vassilevich:2003xt} and impose the Dirichlet condition on half of the spinor space. In odd dimensions the (Hermitian) projector can be constructed by using just the perpendicular gamma matrix (we use the Pauli matrix representation where $-i\gamma_n \equiv  \sigma_3$)
\begin{equation}
\Pi_{\pm}= \frac{1}{2}(\ONE \pm \sigma_3) 
\end{equation}
and Dirichlet condition reads\footnote{In Minkowski signature $(+,-,-)$, the matrix $\chi=\sigma_3$  satisfies $\bar{\chi} = \gamma_0 \sigma_3 \gamma_0= - \bar{\chi} $, which implies the desired properties $\bar{\chi}\gamma_i = \gamma_i \chi$, $\bar{\chi}\gamma_n = -\gamma_n \chi$  of \cite{McAvity:1993ue, Vassilevich:2003xt}. The second half of the spinor space satisfies Neumann condition $\partial_n \Pi_{\pm} \psi |_{x^n=0}=0$.}
\begin{equation} \label{eq:FermBC}
\Pi_{\mp} \psi |_{x^n=0}= 0  ; \quad  \bar{\psi}  \Pi_{\pm}|_{x^n=0} = 0 \quad .
\end{equation}
The above equation therefore defines two different boundary conditions that  project away either left/right handed component of the $2d$ Weyl fermion at the boundary. Note that it can be readily verified that \eqref{eq:FermBC} implies \eqref{eq:CurrBC2} and \eqref{eq:Cardy}.
The boundary condition \eqref{eq:FermBC} also imposes the form of the fermionic Green's function
\begin{equation} \label{eq:FermPropagator}
\vev{\psi(x) \bar{\psi}(x')}_{\pm}= C_d \left(\slashed \partial_x \frac{1}{|x-x'|^{(d-2)}} + \chi \slashed \partial_{\bar{x}} \frac{1}{|\bar{x}-x'|^{(d-2)}} \right) \quad ,
\end{equation}
where $\bar{x}= (x^i,-x^n)$ is the position of the image, $C_d= \frac{\Gamma(\frac{d}{2})}{2 (d-2) \pi^{\frac{d}{2}}}$ and the boundary conditions \eqref{eq:FermBC} are encoded in $\chi= \Pi_+ - \Pi_-= \pm \sigma_3$.
By writing the above function in this form we can readily define Feynman rules analogical to the scalar ones used in \cite{Prochazka:2018bpb}. The free propagators being replaced by fermionic ones $\frac{\slashed p}{p^2}$ and $\frac{\tilde{\slashed p}}{p^2}$ respectively and the 'vertex' $\chi=\pm \sigma_3$ (cf. Figure \ref{fig:P1}).
Let us now briefly discuss how to perform perturbative calculations in this set up (for more details we refer the reader to the Section 3.2 of \cite{Prochazka:2018bpb}).\footnote{A similar perturbative approach together with a more detailed discussion of boundary momentum space was given in an earlier work \cite{Bajnok:2003dj}.}
Momentum space computations work as usual with standard propagators replaced by the ones on Figure \ref{fig:P1}. The dashed line on the reflected propagator should be treated as an external line absorbing the momentum $(0,0,2p_n)$ (even if it is attached to an internal propagator). Vertices are treated as usual - conserving all the components of momentum and the consistency is achieved by  including an overall delta function $\delta(\sum_i (p_n)_i)$, where the sum over normal momenta includes physical external momenta as well as virtual reflected momenta of dashed lines.\footnote{The dashed lines can also be attached to the internal propagators.} \\

\subsection{U(1) anomaly computation}
\begin{figure}[h]
\begin{center}
  \includegraphics[width=80mm]{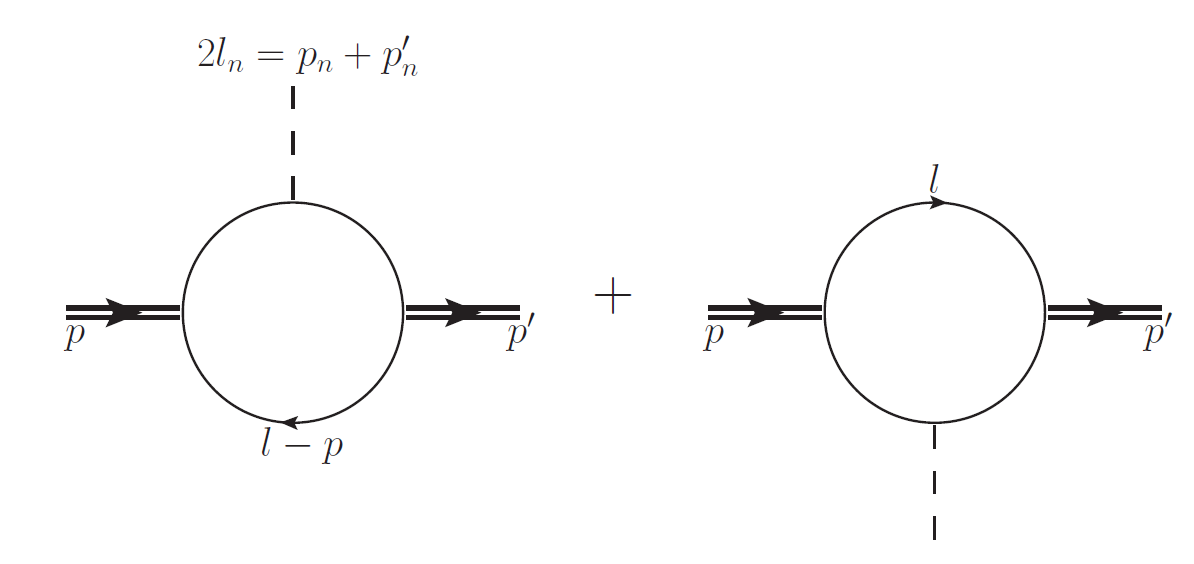}
  \caption{\small The two diagrams contributing to the anomalous part of $\vev{jj}, \vev{TT}$ correlators. Double lines at the edges represent operator insertions with relevant vertices defined in \eqref{eq:Jvertex}, \eqref{eq:Tvertex}. The perpendicular loop momentum is fixed via $\delta(-2l_n+p_n+p_n')$.}
  \label{fig:P2}
  \end{center}
\end{figure}

In this section we will compute the quantities of interest $\Pi^{ij}, \Pi^{in}$ (cf. \eqref{eq:JJmomRep}) for the free fermion example at hand.
To proceed with the computation we can use the usual momentum space methods and write down the usual $\bar{\psi} \psi j_\mu$ vertex corresponding to the insertion of \eqref{eq:FreeFermJ} in the  momentum space
\begin{equation} \label{eq:Jvertex}
V_\mu(p, q_1,q_2) = i \gamma_\mu \; ,
\end{equation} 
where $q_1 , q_2$ are the momenta of incoming fermions satisfying $p=q_1+q_2$. Next we proceed to compute the one-loop diagrams contributing to \eqref{eq:JJmomRep} using the propagator on Figure \ref{fig:P1}. There will be $4$ possible one-loop diagrams contributing to the $\vev{jj}$ correlators, however only the diagrams on Figure \ref{fig:P2} will contribute. This is because the other two diagrams with zero and two dashed outgoing lines are proportional to $\delta(p_n)$ and therefore vanish for $p_n>0$.
These diagrams can now be computed after imposing the momentum conserving delta function $\delta(2l_n-p_n)$ which turns the respective integral into a massive $2d$ bubble with effective mass $m=\frac{p_n}{2}$ (some details of how these computations work are given in the Appendix B of \cite{Prochazka:2018bpb}). Note that the fermionic trace has to be evaluated with the insertion of $\chi= \pm \sigma_3$ at the dashed line vertex so the results for two boundary conditions in \eqref{eq:FermBC} differ by a minus sign. We computed the relevant integrals and traces using 't Hooft-Veltman formalism \cite{tHooft:1972tcz} and verified that the divergent parts cancel out leaving us with a finite result
\begin{eqnarray}\label{eq:CurrExample} \notag
\Pi^{ij}(\underline{p},p_n)&=& \pm \frac{i}{2\pi} \left[\frac{\epsilon^{ik}p_k p^j + \epsilon^{jk}p_k p^i}{\underline{p}^2} \left(1-a(\frac{\underline{p}^2}{p_n^2})\right)+ \epsilon_{ij} a(\frac{\underline{p}^2}{p_n^2})\right] \; , \\
\Pi^{in}(\underline{p}, p_n)&=& \pm \frac{i}{\pi} a(\frac{\underline{p}^2}{p_n^2})\epsilon_{ij} p^j \; ,
\end{eqnarray}
where $a(x^2)= -\frac{1}{2\sqrt{x^2(x^2-1)}} \ln \left(\frac{1+\sqrt{\frac{x^2}{x^2-1}}}{1-\sqrt{\frac{x^2}{x^2-1}}}\right)$. Notice that \eqref{eq:CurrExample} is exactly of the form \eqref{eq:CurrWIgenSol} since the parity-even, non-anomalous part doesn't contribute to the diagrams on Figure \ref{fig:P2}. Furthermore we have 
\begin{eqnarray} \notag
\lim_{x^2\to \infty} a(x^2)&=&0 \; ,\\
\lim_{x^2\to 0} a(x^2)&=&1 \; ,
\end{eqnarray}
which is exactly what we expect from \eqref{eq:FlowP} and \eqref{eq:FhighP}. For the intermediate momenta, the correlator \eqref{eq:CurrExample} is manifestly non-local.\footnote{Note that the branch point at $x^2=1$ can be rewritten in terms of the full Minkowski 3-momentum as $x^2-1= \frac{p_\mu p^\mu}{p_n^2} = 0$. It therefore corresponds to a bulk IR divergence $p^2 \to 0$. We can avoid it by choosing to work on the principal branch where $a$ is real and finite.} \\
By comparing \eqref{eq:CurrExample} and \eqref{eq:CurrWIgenSol} we can extract the value of the anomaly for two different boundary conditions \eqref{eq:FermBC}
\begin{equation} \label{eq:JanomRes}
a_\partial^\pm= \pm \frac{1}{2} \; .
\end{equation}
Let us first observe that the fractional part of this anomaly exactly agrees with the result of \cite{Closset:2012vp}, where it was explained that the half-integer value also implies the parity anomaly in the bulk. On the other hand by taking the $p_n \to 0$ limit of \eqref{eq:CurrExample} we see that fluctuations parallel to the boundary include an 'emergent' chiral current with fractional charge. The anomaly \eqref{eq:JanomRes} cannot be canceled by adding boundary degrees of freedom as these would violate the Dirac quantization condition. Instead it could be canceled by bulk degrees of freedom, namely the $U(1)_{2}$ bulk TQFT as mentioned at the end of Section \ref{sec:AnCanc}. \\
 For even number of massless bulk fermions, the anomaly is an integer and we could cancel it by chiral fermions at the boundary which was also described in Section \ref{sec:AnCanc}. \\    
The integer part of \eqref{eq:JanomRes} can be checked by comparing with the massive fermion computations in the literature.
Recently a contribution to anomaly from integrating out a massive fermion was evaluated in \cite{Dimofte:2017tpi}, where it was found that independently of the sign of the mass one obtains
\begin{equation} \label{eq:delJanom}
\Delta a_\partial^\pm= \pm \frac{1}{2} \; .
\end{equation}
The total IR anomaly of the massive theory is obtained by adding \eqref{eq:delJanom} to \eqref{eq:JanomRes}, which is to be understood as the UV anomaly of the massless theory. Therefore we find
\begin{equation}
{a_\partial^\pm}_{\text{massive}}=a_\partial^\pm+\Delta a_\partial^\pm= \pm 1 \; ,
\end{equation}
which is exactly the value obtained in \cite{Aitken:2017nfd} by introducing Pauli-Villars fields.

\subsection{Gravitational anomaly computation}
In this section we will present the outcome of the calculation of $A^{ijkl},A^{inkl}$ (cf. \eqref{eq:TTmomRep}) analogous to the results in the previous section.  
A $\bar{\psi} \psi T_{\mu \nu}$ vertex corresponding to the insertion of \eqref{eq:FreeFermT} in momentum space reads \cite{Hathrell:1981gz}
\begin{equation}\label{eq:Tvertex}
V_{\mu \nu}(p, q_1,q_2)= \frac{1}{4} \left( \gamma_\mu (q_1-q_2)_\nu+\gamma_\mu (q_1-q_2)_\nu - 2 \eta_{\mu \nu} (\slashed q_1 - \slashed q_2)\right) \quad ,
\end{equation}
where $p=q_1+q_2$.
The computation is identical to the one in previous section with vertices on Figure \ref{fig:P2} replaced by \eqref{eq:Tvertex}. The result reads
\begin{eqnarray} \label{eq:Tenexample} \notag
A^{ij kl}(\underline{p},p_n)&=&  \pm \frac{i}{24 \pi}[1-b(\frac{\underline{p}^2}{p_n^2})]\left( \frac{(\epsilon^{im}p_m p^j+ \epsilon^{jm}p_m p^i)P^{kl} + (\epsilon^{km}p_m p^l+ \epsilon^{lm}p_m p^k)P^{ij}}{\underline{p}^2}\right) \\ \notag
& & + \frac{i}{48 \pi }b(\frac{\underline{p}^2}{p_n^2})\left((\epsilon^{ik}P_{jl}+(i \leftrightarrow j)) + (k \leftrightarrow l) \right)\\
A_{an.}^{in kl}(\underline{p}, 0)&=&\mp\frac{1}{p_n} \frac{i}{12 \pi } b(\frac{\underline{p}^2}{p_n^2}) P^{kl}  \epsilon^{ij} p_i\; ,
\end{eqnarray}
where 
\begin{equation}
b(x^2)=\frac{3}{x^2} \left( (1-x^2)a(x^2)-1 \right) \; ,
\end{equation}
where the function $a$ was defined below \eqref{eq:CurrExample} .
Note again, that the function $b$ has the expected behaviour
\begin{eqnarray} \notag
\lim_{x^2\to \infty} b(x^2)&=&0 \; ,\\
\lim_{x^2\to 0} b(x^2)&=&1 \; .
\end{eqnarray}
By comparing \eqref{eq:Tenexample} and \eqref{eq:TenWIgenSol} we obtain the anomaly
\begin{equation} \label{eq:TanomResD}
e_\partial^{\pm}= \pm 1 \; ,
\end{equation}
whose integer character again agrees with the bulk results of \cite{Closset:2012vp,Kurkov:2018pjw}.
Note that above computation is valid for a Dirac fermion. 
Had we instead considered a Majorana fermion in computing \eqref{eq:Tenexample} we would get an extra $\frac{1}{2}$ from the symmetry factor of diagrams on Figure \ref{fig:P2}. This means that the respective anomaly would be half of the Dirac one
\begin{equation} \label{eq:TanomRes}
{e_\partial^\pm}_{(\text{Majorana})}= \pm \frac{1}{2} \; .
\end{equation}
Here we find agreement with \eqref{eq:Tcondition}
\begin{equation}
{e_\partial^+}_{(\text{Majorana})}- {e_\partial^-}_{(\text{Majorana})}=1 \in \mathbb{Z} \; .
\end{equation}
Before we conclude this section let us remark our direct calculation also shows that the parity-even, conserved contribution to \eqref{eq:Tenexample} vanishes. In \cite{Prochazka:2018bpb} the parity-even part of $A_{ijkl}$ was related to the boundary conformal anomaly proportional to $\hat{R}$, which therefore has to vanish for this example. Indeed this is what was found by a direct heat kernel computation in \cite{Fursaev:2016inw}.

\section{Discussion and conclusions} \label{sec:Conclusions}
In this paper we studied the consequences of anomalies \eqref{eq:currAn}, \eqref{eq:tenAn} for the two-point functions of bulk operators $j^\mu$, $T^{\mu \nu}$ subject to boundary conditions \eqref{eq:CurrBC2}, \eqref{eq:Cardy} respectively. 
As a result we obtained the anomalous solutions \eqref{eq:CurrWIgenSol}, \eqref{eq:TenWIgenSol} which relate the anomalies to flat space correlators. In this is sense we extend the known results \cite{McAvity:1995zd}, which only consider non-anomalous, parity even solutions. The results of Section \ref{sec:Momentum} should hold in a generic bCFT subject to \eqref{eq:CurrBC2} and \eqref{eq:Cardy}, although some care has to be taken when the theory includes decoupled boundary degrees of freedom (cf. the discussion under \eqref{eq:Jcondition}). In particular, we have shown that the coefficient of the continuous solution satisfying \eqref{eq:FlowP}, \eqref{eq:FhighP} is a physical quantity (and an analogous conclusion follows for the gravitational solution in Section \ref{sec:EMTcorr} ). \\
To convince the reader of existence of such anomalous terms and their computability we calculated the relevant correlators \eqref{eq:JJmomRep}, \eqref{eq:TTmomRep} directly in momentum space using the technique introduced in \cite{Prochazka:2018bpb}. The computation revealed that terms of the form \eqref{eq:CurrWIgenSol}, \eqref{eq:TenWIgenSol} indeed appear and we extracted the respective anomalies \eqref{eq:JanomRes}, \eqref{eq:TanomResD},  \eqref{eq:TanomRes} from them. \\
The momentum space expression \eqref{eq:CurrWIgenSol} (and \eqref{eq:TenWIgenSol} by extension) has a natural physical interpretation. The $p_n \to 0$ limit \eqref{eq:HighMomAn} describes states that never leave the boundary and therefore only experience the anomaly through two-dimensional boundary currents. The states with $p_n>0$, on the other hand, are allowed penetrate into the bulk where they start experiencing bulk CS currents so the anomaly appears through a linear combination of both phenomena, which is indeed what the solution \eqref{eq:CurrWIgenSol} describes.\\
Furthermore, given the non-trivial dependence on $p_n$ (see \eqref{eq:CurrExample} and below) the solutions \eqref{eq:CurrWIgenSol}, \eqref{eq:TenWIgenSol} have to arise from \textit{non-local} bulk terms in the effective action in a generic bCFT. In summary we see that in general it is insufficient to consider just the CS terms in the effective action to describe the correct physics in the presence of a boundary.
 To some extent this observation is similar to the known subtlety \cite{Seiberg:2016gmd} that the parity-odd contribution to the path integral of three-dimensional massless fermion is actually given by the $\eta$-invariant rather than $\frac{1}{2}$ times the CS term as is sometimes assumed. \\
 Our use of momentum space techniques and the calculation methods of Section \ref{sec:FreeFerm} are quite versatile and can be readily applied to study other anomalies. For example we checked explicitly vanishing of the mixed gauge-gravitational anomaly of free massless fermion. An interesting extension of this work  would be to apply the present momentum space formalism to constrain the form of potential parity-odd trace anomalies in a generic bCFT. 
 \\
 We also found that the dependence of $a_\partial$, $e_\partial$ upon boundary conditions is restricted by \eqref{eq:Jcondition}, \eqref{eq:Tcondition}. In particular in theories with boundary RG flows\footnote{By this we understand flows with nonzero trace of $\tilde{t}_{ij}$ in \eqref{eq:bTensor}. Such flow can be interpreted as a flow of boundary conditions under a deformation by relevant operators at the boundary.} satisfying  \eqref{eq:CurrBC2}, \eqref{eq:Cardy} at the ends of the flow, these conditions could provide extra constraints on the allowed asymptotics.  \\
 Finally, we would like to remark that the connection between parity anomaly in (2+1) dimensions and quantum Hall effect with $\nu=\pm \frac{1}{2}$ (cf. \cite{Jackiw:1984ji,Haldane:1988zza}) manifests itself through \eqref{eq:CurrExample} and \eqref{eq:JanomRes}. Indeed the correlator \eqref{eq:CurrExample} implies anomalous currents with half-integer Hall conductance running parallel to the boundary and parity-violating CS term in the bulk. The results of this paper have therefore potential to provide a momentum space perspective on this and other similar boundary phenomena in condensed matter physics.

\subsection*{Acknowledgements}

The author is thankful to Zohar Komargodski for commenting on the manuscript and Guido Festuccia for productive discussions on the topic of the paper. Additionally the author thanks Arash Ardehali, Zoltán Bajnok,  Kostas Skenderis and Alexander Söderberg for stimulating and relevant conversations. 
The author is supported by the ERC STG grant 639220 (curvedsusy). 

\newpage

\bibliographystyle{utphys}
\bibliography{CPbdary}

\end{document}